# Ultrafast and electrically tunable Rabi frequency in a germanium hut wire hole spin qubit

He Liu,[1,2,#] Ke Wang,[1,2,#] Fei Gao,[7,3] Jin Leng,[1,2] Yang Liu,[1,2] Yu-Chen Zhou,[1,2] Gang Cao,[1,2,8] Ting Wang,[3,8] Jianjun Zhang,[3,8] Peihao Huang,[4,5,6,*] Hai-Ou Li,[1,2,8,*] and Guo-Ping Guo[1,2,8,9,*]

[1] *CAS Key Laboratory of Quantum Information, University of Science and Technology of China, Hefei, Anhui 230026, China*

[2] *CAS Center for Excellence and Synergetic Innovation Center in Quantum Information and Quantum Physics, University of Science and Technology of China, Hefei, Anhui 230026, China*

[3] *Institute of Physics and CAS Center for Excellence in Topological Quantum Computation, Chinese Academy of Sciences, Beijing 100190, China*

[4] *Shenzhen Institute for Quantum Science and Engineering, Southern University of Science and Technology, Shenzhen 518055, China*

[5] *International Quantum Academy, Shenzhen 518048, China.*

[6] *Guangdong Provincial Key Laboratory of Quantum Science and Engineering, Southern University of Science and Technology, Shenzhen, 518055, China*

[7] *Qilu Institute of Technology, Jinan, 250200, China*

[8] *Hefei National Laboratory, University of Science and Technology of China, Hefei 230088, China*

[9] *Origin Quantum Computing Company Limited, Hefei, Anhui 230026, China*

[#]These authors contributed equally to this work

[*]Corresponding authors: huangph@sustech.edu.cn; haiouli@ustc.edu.cn; gpguo@ustc.edu.cn;

## Abstract

Hole spin qubits based on germanium (Ge) have strong tunable spin orbit interaction (SOI) and ultrafast qubit operation speed. Here we report that the Rabi frequency ($f_\text{Rabi}$) of a hole spin qubit in a Ge hut wire (HW) double quantum dot (DQD) is electrically tuned through the detuning energy ($\epsilon$) and middle gate voltage ($V_\text{M}$). $f_\text{Rabi}$ gradually decreases with increasing $\epsilon$; on the contrary, $f_\text{Rabi}$ is positively correlated with $V_\text{M}$. We attribute our results to the change of electric field on SOI and the contribution of the excited state in quantum dots to $f_\text{Rabi}$. We further demonstrate an ultrafast $f_\text{Rabi}$ exceeding 1.2 GHz, which evidences the strong SOI in our device. The discovery of an ultrafast and electrically tunable $f_\text{Rabi}$ in a hole spin qubit has potential applications in semiconductor quantum computing.

Spin qubits based on semiconductor materials are being employed as building blocks for scalable quantum computing.[1,2] Substantial progress has been made based on silicon quantum dots in recent years due to their superior coherence time and compatibility with CMOS techniques.[3-24] Recently, researchers have intensively explored hole spin-qubit systems and experimentally realized single- and two-qubit gates and four-qubit entanglement.[13, 14, 25-29] Compared to electrons, the p-type symmetry of the valence band of holes excludes the contact hyperfine interaction, which extends the coherence time.[30-33] Moreover, the strong SOI of holes[32, 34-36] allows all electrically controlled electric-dipole spin resonance (EDSR) without fabricating a micromagnet to generate artificial SOI,[9, 11, 17] which simplifies the fabrication process of spin qubits. The Rabi frequency $f_{\text{Rabi}}$ is an important parameter that determines the properties of a spin qubit. Some factors can affect $f_{\text{Rabi}}$, such as $g$-tensor modulation,[37-42] $g$ factor difference between two quantum dots,[43] and the strength of the SOI.[28] However, $g$ tensor modulation and $g$ factor difference will make a significant contribution to $f_{\text{Rabi}}$ only when the direction of the applied magnetic field is not perpendicular to the sample.[42,43] In this work, we apply a perpendicular magnetic field to the sample, the contribution of the above two factors to $f_{\text{Rabi}}$ is less significant compare to the contribution of spin-orbit interaction mechanism. Therefore, we are more interested in the effect of SOI strength on $f_{\text{Rabi}}$. For instance, the $f_{\text{Rabi}}$ of hole spin is generally 1 ~ 2 orders of magnitude higher than that of electron spin due to the much stronger intrinsic SOI.[13, 25-29] Compared with the GaAs hole system[44, 45], the one-dimensional Ge HW hole system has stronger direct Rashba SOI, which allows the spin qubit in this system has a higher $f_{\text{Rabi}}$ upper limit.[35] Theoretical analyses reveal that in 1D and 2D Ge materials, the external electrostatic field modulates $f_{\text{Rabi}}$ by changing the SOI.[35, 36, 46-48]

In nanowire systems, researchers have previously achieved electrically tunable $f_{\text{Rabi}}$ in a Ge/Si core/shell nanowire[28] and ultrafast $f_{\text{Rabi}}$ in a Ge HW[29] and a Ge/Si core/shell nanowire.[28] To obtain higher gate fidelity for applications such as quantum error correction, it is desirable to further optimize the speed of the EDSR for fast electric manipulation of a hole spin. Here, we explore the electrical tunability and a record of $f_{\text{Rabi}}$ in our Ge HW system. The $f_{\text{Rabi}}$ is tunable by changing the detuning energy ($\epsilon$) and middle gate voltage ($V_{\text{M}}$) of the DQD. To dig in our results, we establish a hole EDSR model. In addition to the contribution of SOI to $f_{\text{Rabi}}$, the detuning $\epsilon$ and excited state also affect the strength of $f_{\text{Rabi}}$. Through our analysis, we can conclude that the difference in the electric field caused by changing $\epsilon$ and $V_{\text{M}}$ can affect the intensity of the SOI to modulate $f_{\text{Rabi}}$. Then, we further show an ultrafast $f_{\text{Rabi}} >$ 1.2 GHz of the EDSR of a hole spin with a qubit frequency of approximately 3.75 GHz, representing the speediest quantum operation of a single spin state to our knowledge. The subnanosecond $1/f_{\text{Rabi}}$ manifests the strong SOI in our device. It also provides essential guidance for further optimization of the gate fidelity of a hole spin qubit for fault-tolerant quantum computing applications.

Figure 1a shows a false-colored scanning electron microscope image of the DQD device based on a Ge HW. We have performed universal coherent control of a hole spin qubit in a similar device.[29, 49] Two source (S) and drain (D) electrodes are processed on the nanowire to form ohmic contacts. Three top gates L, M, and R with direct current (DC) voltage are used to confine the DQD and tune the energy level of each quantum dot. A 25-nm-thick $Al_2O_3$ insulating layer between the Ge HW and top gates is used to prevent current leakage. In particular, microwave bursts together with pulse sequences are applied to gate R through a bias tee to provide high-frequency electric control. Experimentally, the state signals are obtained by DC current from transport measurement. We tune the voltage of the three top gates to obtain the "current triangle" in the DQD, as shown in Figure 1b, with a 3 mV DC bias voltage applied to the source and drain. The Pauli spin blockade (PSB) region is surrounded by blue and black dashed lines in Figure 1b. In this region, the leakage current is suppressed because the carrier transport is blocked due to the existence of the triplet spin state. While the true charge transition in the PSB region in Figure 1b is between the (m,n+2) and (m+1,n+1) charge states, we can simplify it as a (0,2) to (1,1) charge transition when the total hole spin of each dot is zero in the (m,n) state.

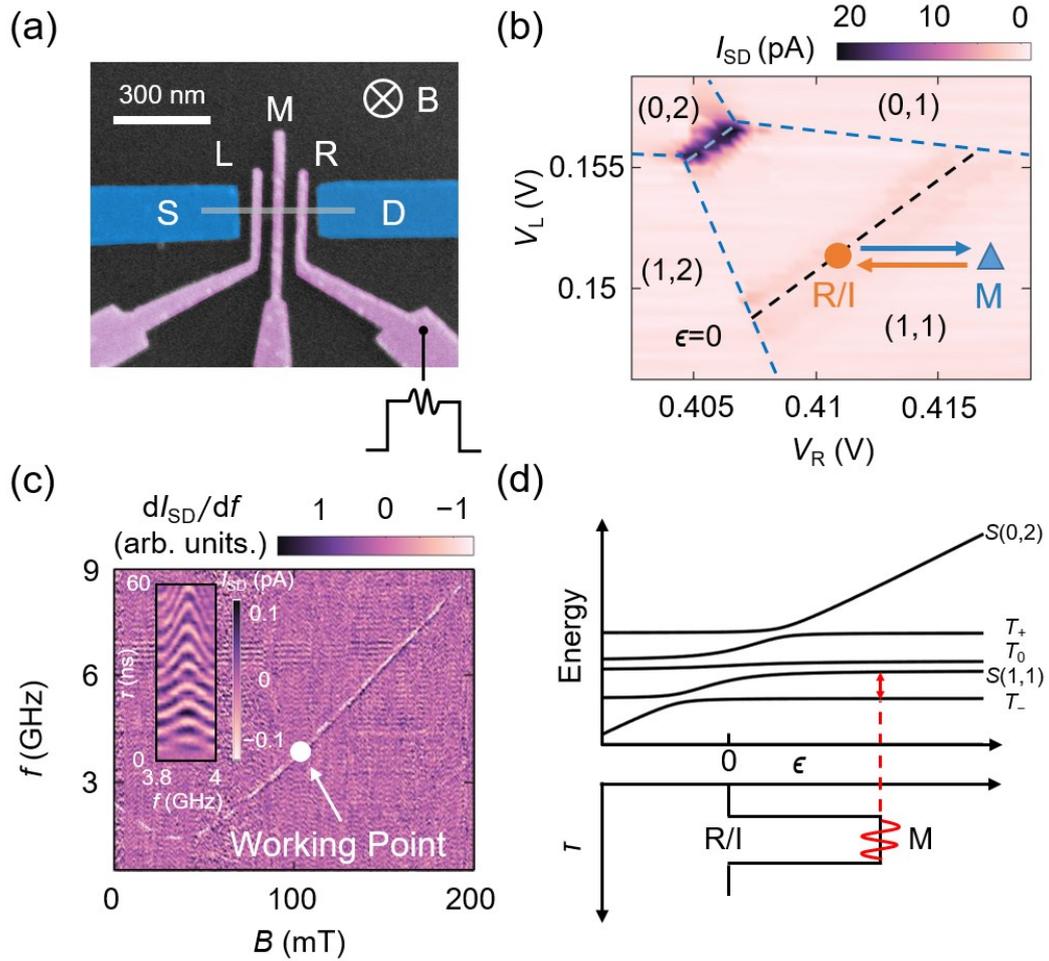

**Figure 1.** Experimental setup and definition of a hole spin qubit. (a) Scanning electron microscopy image of the hole DQD in a Ge HW. The structure consists of source (S), drain (D), and three top gates (L, M, R). To lift the spin state degeneracy, we apply a static magnetic field vertical to the device to Zeeman split the spin states. A high-frequency microwave burst is applied to gate R to manipulate the state of the spin qubit in the DQD. (b) Measurement of a bias triangle under PSB conditions. We marked the charge transitions of the triangle by blue dashed lines and the equivalent charge states. The black dashed line indicates the zero detuning line, and the PSB region is surrounded by blue and black dashed lines. During the experiment, we measured the transport current through the nanowire to distinguish the spin state. The manipulation of spin qubits consists of three stages: initialization (I), manipulation (M) and readout (R). The orange circle represents the position of state R/I in the PSB region when $\epsilon = 0$, and the blue triangle represents the position of state M in the CB region. These stages can

4/19

be switched by applying a pulse sequence on gate R. (c) EDSR spectrum of a hole spin qubit. An EDSR line is shown in this figure, and we choose $f = 3.75$ GHz as the working point of the spin qubit. In the PSB region, we apply a continuous microwave burst with a power $P = -15$ dBm. When the microwave frequency resonates with the frequency of the spin qubit, spin flip occurs. With the lifting of PSB, we observe an increase in the transport current. The Larmor frequency of the spin qubit changes with the applied magnetic field, and we can fit this EDSR line as a $|(1,1)S\rangle$ to $|(1,1)T_-\rangle$ transition. The white dashed line shows our calculation results for this EDSR line. The inset of this figure shows the coherent control of a hole spin qubit. At a microwave driving power $P = 0$ dBm, we measure the leakage current as a function of microwave burst duration time ($\tau$) and microwave driving frequency ($f$), which is called Rabi oscillation. (d) Top: Energy of the five relevant spin eigenstates in the DQD as a function of detuning. Bottom: The pulse sequence used to switch between stage R/I and stage M.

To achieve qubit operation, we apply a magnetic field ($B$) perpendicular to the sample plane to split the Zeeman energy in the PSB region. As presented in Figure 1a, a continuous microwave burst is applied to manipulate the spin state via gate R.[29] When the microwave energy resonates with the spin splitting energy, the spin state flips periodically due to EDSR. We write the Hamiltonian of this DQD system on the basis of the lowest few two-hole states, $|(0,2)S\rangle$, $|(1,1)S\rangle$, $|(1,1)T_0\rangle$, $|(1,1)T_+\rangle$, $|(1,1)T_-\rangle$. We fit this EDSR signal with the transition between $|(1,1)S\rangle$ and $|(1,1)T_-\rangle$, and the calculated fitting of this EDSR transition is shown by white dashed lines in Figure 1c. The specific calculations are shown in Section S2 of the supporting information. A linear correlation between the resonant frequency $f$ and the value of $B$ can be extracted in the range of 90 ~ 200 mT.

To coherently control a hole spin qubit, we generate modulated waveforms to achieve Rabi oscillation. The operation process is divided into three stages: initialization (I), manipulation (M) and readout (R). The positions of the three operation points in the charge stability diagram are shown in Figure 1b with orange circles and blue triangles. At the initialization stage, the spin state is in the PSB regime, and the charge occupation is in the equivalent (1,1) state. The transport current is suppressed due to PSB, which is used to distinguish the spin state. At the manipulation stage, we pulse the charge state to the Coulomb blockade (CB) regime and apply a microwave burst resonating with the frequency of the spin qubit. In this stage, the spin state rotates coherently controlled by the microwave burst. At the readout stage, we pulse the charge state back to the PSB regime for spin state readout. The periodic flip of the spin state can be reflected by the oscillations of the transport current, and the Rabi chevron is shown in the inset of Figure 1c. The frequency with the largest fringe interval in the Rabi chevron corresponds to the center frequency of the qubit ($f_{qubit}$). We can scan a single Rabi oscillation curve at $f = f_{qubit}$, and extract $f_{Rabi}$ according to its oscillation period. In order to better describe the above physical processes, we show the energy of the five relevant spin eigenstates in the DQD as a function of detuning at the top of Figure 1d. The pulse in the bottom of Figure 1d is used to switch between stage R/I and stage M.

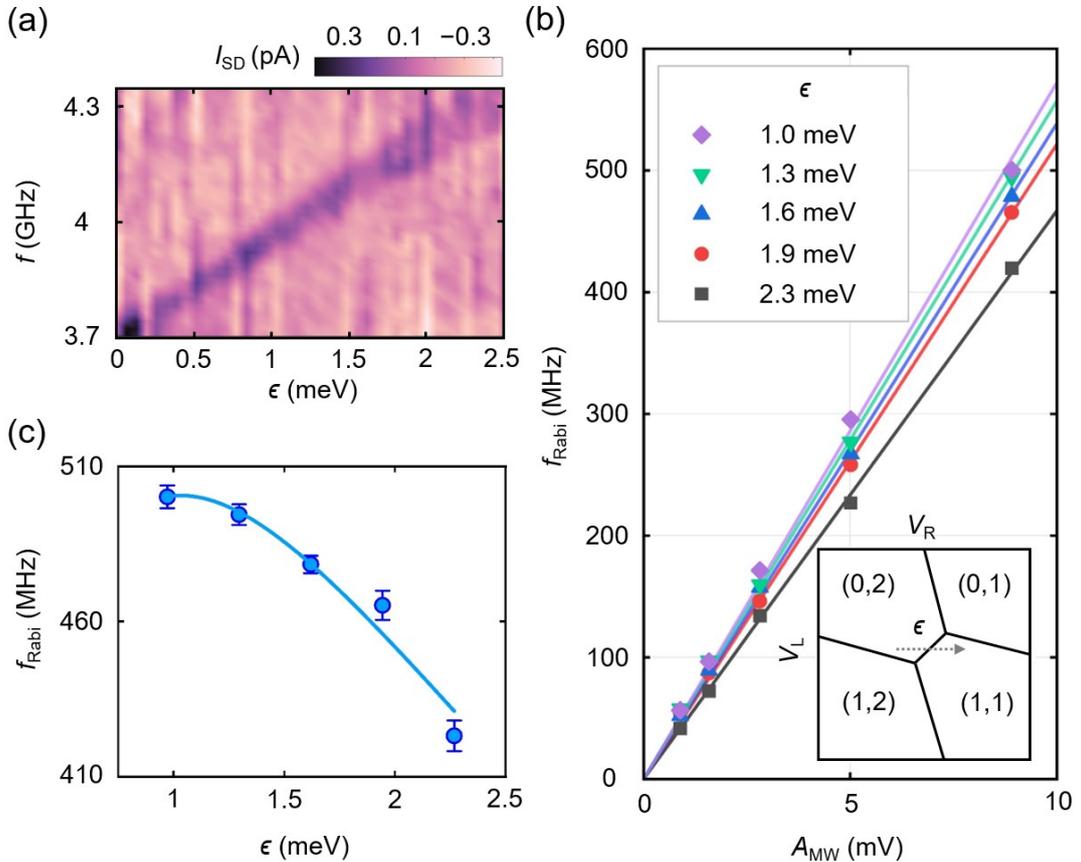

**Figure 2.** Detuning energy tunability of qubit properties. (a) Qubit frequency as a function of $\epsilon$, $B$ is fixed at 100 mT. The pulse amplitudes are converted to different detuning energies. (b) $f_{\text{Rabi}}$ dependence of the amplitude of microwave under different $\epsilon$. The gray arrow in the inset indicates the direction in which $\epsilon$ becomes larger. We measure the data at $V_M = 27.5$ mV. We obtain $f_{\text{Rabi}}$ from $P = -15$ dBm to $P = 5$ dBm. $A_{\text{MW}}$ is the amplitude of microwave, the relationship between $A_{\text{MW}}$ and $P$ is: $A_{\text{MW}}(\text{mV}) = (2 * P(\text{mW}) * 50\Omega)^{1/2}$, $P(\text{mW}) = 10^{(P(\text{dBm})-36/10)}$, where 36 dBm is the attenuation of microwave in the transmission process. $f_{\text{Rabi}}$ is linearly proportional to the amplitude of the driving power. The five sets of data under different $\epsilon$ are slightly different, but they have the same trend with respect to the change in the driving power. (c) $f_{\text{Rabi}}$ as a function of $\epsilon$. In order to control the variables and simplify the process of the related analysis, we keep a constant qubit frequency as $f_{\text{qubit}} = 3.75$ GHz by changing the magnetic field at different $\epsilon$ during the measurement. To show the changing characteristics of the data more clearly, we show $f_{\text{Rabi}}$ as a function of $\epsilon$ under $P = 5$ dBm. $f_{\text{Rabi}}$ decreases gradually with increasing $\epsilon$. The blue solid line represents the fitting results by eq 2.

As we have mentioned above, several theoretical analyses have revealed that an electric field can modulate the SOI to change the $f_{\text{Rabi}}$ of a hole spin qubit.[35,46] The dependence of $f_{\text{Rabi}}$ on gate voltages has also been shown in some experimental works.[28,29] However, there is no comprehensive study on the influence of gate voltages on the $f_{\text{Rabi}}$ of spin qubits in an HW system, and a detailed theoretical analysis is still lacking. The hole system is a relatively complex system, and physical quantities, such as the SOI and the spin-orbital states, could vary and affect $f_{\text{Rabi}}$. It is an important and challenging task to understand the specific dependency relation between $f_{\text{Rabi}}$ and the tunable variables. During our measurement of a hole spin qubit, the electrically tunable quantities are the detuning energy ($\epsilon$) and middle gate voltage ($V_M$). In the following, we investigate the relation between $f_{\text{Rabi}}$ and these quantities and the tunability of $f_{\text{Rabi}}$.

First, we study the relative results of spin qubit properties arising from the change in $\epsilon$, which can be tuned by applying square pulses with varying amplitudes to gate R during the measurement of spin qubits. We have measured the lever arm of electrode R relative to the source-drain bias energy in Section S1 of the supporting information, and thus, we can convert the pulse amplitude to $\epsilon$ accurately. In previous experiments, the pulse amplitude of the waveform for coherent control was set up arbitrarily, which may not be the best choice. This means that the quality of our qubit could potentially be further optimized by adjusting the pulse amplitude.

As shown in Figure 2a, we keep $B = 100$ mT, and then we extract the qubit frequency dependence of the pulse amplitude, which is converted into $\epsilon$ for clarity. Figure 2b shows the relationship between $f_{\text{Rabi}}$ and microwave driving power ($P$) under different $\epsilon$. To exclude the influence of qubit frequency on $f_{\text{Rabi}}$, along with the increase in $\epsilon$ during our experiment, we change the magnetic field to fix the qubit frequency at $f_{\text{qubit}} = 3.75$ GHz. We plot a schematic diagram of charging stability in the inset of Figure 2b, where the gray arrow indicates the direction of $\epsilon$ increasing. In our results, $f_{\text{Rabi}}$ is proportional to the amplitude of the microwave power at different $\epsilon$. Meanwhile, we can obtain different $f_{\text{Rabi}}$ with different $\epsilon$. To clearly show the influence of $\epsilon$ on $f_{\text{Rabi}}$, we show that $f_{\text{Rabi}}$ gradually increases and then saturates as $\epsilon$ decreases with an applied power of 5 dBm in Figure 2c. This trend is universal, as shown in the data we measured under different driving powers (Section S4 of the supporting information).

To understand the dependence of $f_{\text{Rabi}}$ on the detuning $\epsilon$, we have established a model for the EDSR of holes in an HW DQD. The theoretical derivation is given in Section S3 of the supporting information, and the Rabi frequency $f_{\text{Rabi}}$ of the hole EDSR due to the SOI in the presence of excited orbital states is

$$h f_{\text{Rabi}} \approx e|E_{\text{ac}}|E_Z \sum_n \frac{l_{\text{dot}}^2 t_{\text{c,on}}^2}{l_{\text{SO}}(\epsilon+\Delta_n)^3} \tag{1}$$

where $e$ is the elementary charge, $E_{\text{ac}}$ is the electric field generated by the microwave burst, $E_Z = h f_{\text{qubit}}$ is the Zeeman energy of the spin qubit, $l_{\text{dot}}$ is the distance between two quantum dots, $l_{\text{SO}}$ is the spin orbit length in the Ge HW, we use the value of $l_{SO}^{-1}$ to reflect the strength of the SOI,[34] and $\Delta_n$ is the energy spacing between the ground state and excited state in a dot, such as the right dot. $t_{c,0n}$ is the tunneling between the ground orbital state in one dot and the $n$-th orbital state in the other. Specifically, when n = 0, $t_{c,00}$ is the tunnel coupling between the $|(1,1)S\rangle$ and $|(0,2)S\rangle$ ground orbital states, and $\Delta_0 = 0$. $\epsilon$ represents the detuning energy in the DQD.

The variation in detuning energy $\epsilon$ changes the electric field strength, which in turn modifies the SOI. Theoretical work indicates that the strength of the Rashba SOI is proportional to the electric field strength.[35, 46] Thus, with the change in $\epsilon$, we write $1/l_{SO}$ as $\frac{1}{l_{SO}} = b_1 * \Delta\epsilon + \frac{1}{l_{SO}^0}$, where $b_1$ is related to the material properties and $\Delta\epsilon$ is the difference between $\epsilon$ and a fixed detuning in Figure 2c. Suppose the detuning $\epsilon$ is in a regime where the ground orbital in the left dot is the lowest orbital state. Since the tunneling $t_{c,0n}$ between the ground state in the left dot and an excited orbital state $|n\rangle$ in the right dot can be much stronger than the tunneling $t_{c,00}$ between the lowest orbital states in the two dots, $t_{c,0n} \gg t_{c,00} = t_c$,[50, 51] the speed of Rabi oscillation could be enhanced substantially due to the presence of the excited orbital state. In fact, we first considered the influence of the ground orbital states on $f_{\text{Rabi}}$ and found that its contribution was not significant enough to explain our measurement results. Therefore, we have further considered the influence of excited orbital states on $f_{\text{Rabi}}$. In this case, we consider one particular term, i.e., the $n$-th term, in eq 1 with $\Delta_n = E_{ST} = 3$ meV extracted from the "current triangle" in Figure 1b, and neglect other terms. Since the value of $t_{c,0n}$ is unknown, we fit $\frac{t_{c,0n}^2}{l_{SO}^0}$ as a whole, and eq 1 is written as

$$hf_{\text{Rabi}} = e|E_{\text{ac}}|E_Z(b_1 * t_{c,0n}^2 \Delta\epsilon + \frac{t_{c,0n}^2}{l_{SO}^0}) \frac{l_{\text{dot}}^2}{(\epsilon+\Delta_n)^3}. \tag{2}$$

The fitting results are shown in Figure 2c as a blue solid line, where $b_1 * t_{c,0n}^2 = 6.35 * 10^{-2}$ $(1.76 * 10^{-3})$ meV · nm$^{-1}$ and $\frac{t_{c,0n}^2}{l_{SO}^0} = 8.17 * 10^{-2}$ $(1.00 * 10^{-3})$ meV$^2$ · nm$^{-1}$. The specific fitting process and fitting results of $P = -5$ dBm and $P = 0$ dBm are shown in Section S4 of the supporting information. Note that if only the two lowest orbital states in the two dots are considered, one cannot fit the results with realistic values of the parameters. The results of this experiment can be understood as the influence of $\epsilon$ and $l_{SO}$. From eq 2, as $\epsilon$ decreases, despite the strength $1/l_{SO}$ of SOI decreasing, $f_{Rabi}$ is enhanced and saturates at small $\epsilon$, which is consistent with experimental observations. We can also give an estimation of $l_{SO}$. We have extracted $t_c = 8$ ueV at $\epsilon = 0$ from the EDSR spectrum (Section S2 of the supporting information), which $t_c$ is the tunneling between the lowest orbital states in the two dots. According to the theory, the existence of an excited state will cause a significant increase in tunnel coupling $t_{c,0n}$.[51] If we estimate that $t_{c,0n}$ is in the range of 0.1 meV to 0.5 meV, the corresponding maximum value of $l_{SO}$ ranges from 0.1 nm to 3 nm, which is close to the magnitude of $l_{SO}$ extracted from the same sample.[29] From the experiment and analysis, we found that with the increase of $\epsilon$, the orbital hybridization and orbital dipole moments decrease, which leads to the decrease in $f_{Rabi}$. We also emphasize that $l_{SO}$ is tunable by $\epsilon$ because of the change in electric field, so as to affect $f_{Rabi}$. As a result, we can modulate $f_{Rabi}$ by electric field parameters such as $\epsilon$.

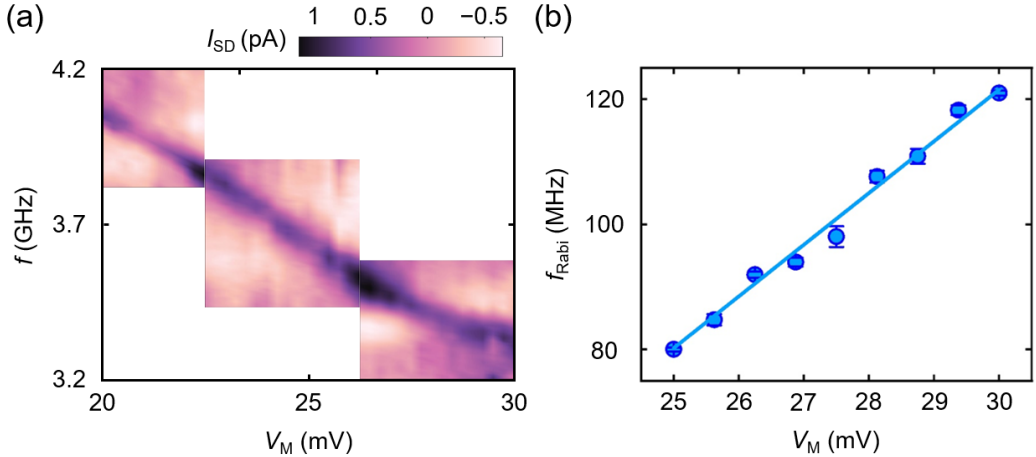

**Figure 3.** Middle gate voltage tunability of qubit properties. (a) Qubit frequency as a function of the voltage of gate M, $B$ is fixed at 100 mT. With increasing $V_M$, the qubit frequency shows a falling trend. (b) $f_{Rabi}$ as a function of $V_M$. At each data point, we fix the qubit frequency at 3.75 GHz to keep the Zeeman energy constant. Considering the data shown in (a), we adjust the magnetic field to compensate for the change in the g-factor. $f_{Rabi}$ is approximately proportional to $V_M$, and $f_{Rabi}$ varies from 80 MHz to 120 MHz. The blue solid line represents the fitting results by eq 3.

We have demonstrated that the difference in the electric field caused by changing $\epsilon$ has a significant impact on the $f_{\text{Rabi}}$ of a hole spin qubit. As a parameter related to the electric field, the change in middle gate voltage $V_M$ should also modulate SOI and affect $f_{\text{Rabi}}$. Next, we study the modulation effect of $V_M$ on $f_{\text{Rabi}}$. To exclude the influence of qubit frequency on $f_{\text{Rabi}}$, we first extract the frequency of the spin qubit as a function of $V_M$, as shown in Figure 3a, $B$ is fixed at 100 mT. Then, the qubit frequency is kept constant at $f_{\text{qubit}} = 3.75 \text{ GHz}$ with different values of $V_M$ by changing the magnetic field. The measurement result is shown in Figure 3b. By changing $V_M$ from 25 mV to 30 mV, $f_{\text{Rabi}}$ increases by nearly 50%, showing that $V_M$ has a high tunability on $f_{\text{Rabi}}$.

Our results in Figure 3b can also be described by eq 1, and we attribute the changes in $f_{\text{Rabi}}$ to the modulation of $V_M$ on $1/l_{\text{SO}}$. Similar to our analysis of the relationship between $\epsilon$ and $1/l_{\text{SO}}$, we write the expression of $1/l_{\text{SO}}$ approximately as $\frac{1}{l_{\text{SO}}} = b_2 * \Delta V_M + \frac{1}{l_{\text{SO}}^0}$, where $b_2$ is related to material properties and $\Delta V_M$ is the difference between $V_M$ and the first measured middle gate voltage in Figure 3b. Then, eq 1 is written as

$$hf_{\text{Rabi}} = e|E_{\text{ac}}|E_Z(b_2 * t_{c,0n}^2 \Delta V_M + \frac{t_{c,0n}^2}{l_{\text{SO}}^0}) \frac{l_{\text{dot}}^2}{(\epsilon + \Delta_n)^3}. \quad (3)$$

The fitting results are shown in Figure 3b as a blue solid line, where $b_2 * t_{c,0n}^2 = 1.22 * 10^{-2} \ (4.94 * 10^{-4}) \text{ meV} \cdot \text{nm}^{-1}$, $\frac{t_{c,0n}^2}{l_{\text{SO}}^0} = 1.15 * 10^{-1} \ (1.47 * 10^{-3}) \text{ meV}^2 \cdot \text{nm}^{-1}$. The specific fitting process is shown in Section S4 of the supporting information. Here, we take $t_{c,0n}$ as a constant value and fit $\frac{t_{c,0n}^2}{l_{\text{SO}}^0}$ as a whole. According to the previous explanation, $t_{c,0n}$ in our system is relatively large. We believe that when $V_M$ changes from 25 mV to 30 mV, the change proportion of $t_{c,0n}$ is small, so ignoring its change will not have a significant impact on the results.

These measurement results reveal that $f_{\text{Rabi}}$ is highly tunable and positively correlated with the increase in electric field strength tuned by $V_M$. The fitting results suggest that the strong modulation of $f_{\text{Rabi}}$ by $V_M$ is due to the increase in $1/l_{\text{SO}}$ with increasing $V_M$. If we estimate $t_{c,0n} = 0.3 \text{ meV}$, $l_{\text{SO}}$ decreases from approximately 0.8 nm to 0.5 nm with the increase in $V_M$ from 25 mV to 30 mV (Section S4 of the supporting information). In Section S5 of the supporting information, we approximately assumed a possible dependency relationship between $t_{c,0n}$ and $V_M$, then we fit the data in Figure 3b and extracted $l_{SO}$. The value and variation trend of $l_{SO}$ are relatively close to the results assuming that $t_{c,0n}$ is a constant value (Section S4 of the supporting information). Therefore, we believe that $l_{SO}$ is the main factor in the modulation of $f_{Rabi}$ by $V_M$.

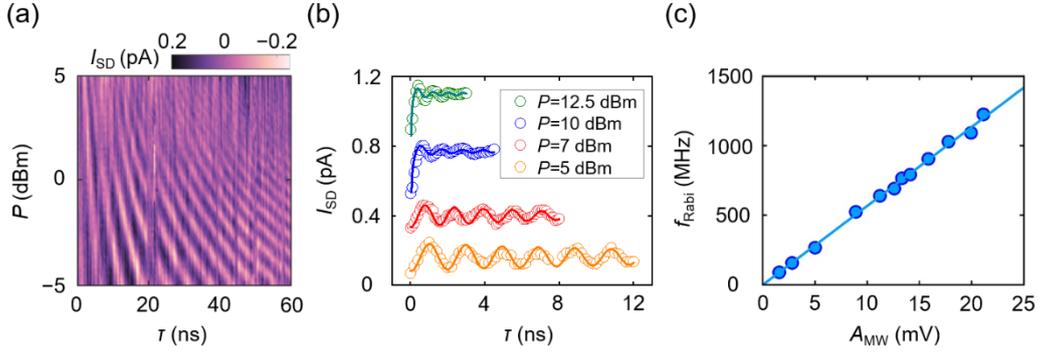

**Figure 4.** Ultrafast hole spin qubit manipulation. (a) Rabi oscillations as a function of driving power ($P$) and microwave burst time ($\tau$). (b) We measure the Rabi oscillations of the spin qubit at the center frequency $f = 3.75$ GHz. The experimental data and fitting results with driving power $P = 5, 7, 10, 12.5$ dBm are shown in this figure. (c) $f_{Rabi}$ dependence of the amplitude of the microwave driving power ($A_{MW}$). $f_{Rabi}$ is linearly proportional to $A_{MW}$ within the measurement range, and the maximum $f_{Rabi}$ exceeds 1.2 GHz.

In Figure 2b, we obtain a $f_{Rabi}$ close to 500 MHz at $P = 5$ dBm, which is a faster operation speed compared to our previous work.[29] This means there is a stronger SOI in our device. To prove this, we are eager to further achieve an even larger $f_{Rabi}$ in this system by increasing the driving power. Figure 4a shows that $f_{Rabi}$ gradually increases with a decreasing interval between stripes while increasing the driving power from $-5$ dBm to 5 dBm. To visualize the results more clearly, we obtain Rabi oscillations and then quantitatively fit $f_{Rabi}$ at different driving powers. Some Rabi oscillations under high driving power are shown in Figure 4b. The Rabi oscillation deteriorates quickly with increasing microwave driving power, which reflects the decrease in decoherence time. This phenomenon is probably because of the heating effect caused by the increase in microwave driving power. The dependence of $f_{\text{Rabi}}$ on the amplitude of the driving power is plotted in Figure 4c, where a linear relationship is clearly depicted. An impressive value of $f_{\text{Rabi}} = 1225 \text{ MHz}$ is obtained at $P = 12.5 \text{ dBm}$, which is amazing since this value is almost one-third of the qubit frequency. The linear relationship between $f_{\text{Rabi}}$ and the amplitude of driving microwaves is maintained during the measurement range of $P$. Therefore, we can speculate that $f_{\text{Rabi}}$ does not reach the upper bound where we should see the saturation. We can also conclude that photon assistant tunneling has no significant impact during the measurement for the dependence between $f_{\text{Rabi}}$ and $A_{\text{MW}}$ does not significantly deviate from linearity in Figure 4c.[29, 52] In summary, we obtain an $f_{\text{Rabi}} > 1.2 \text{GHz}$ with a relatively low qubit frequency. This is quite unusual compared to the results in other experiments based on hole spin qubits and our previous work[26, 28, 29] because our results approach the limitation of the rotating wave approximation.[53] In the future, we will try to operate the spin qubit at higher temperatures.[54] The ultrafast $f_{\text{Rabi}}$ may improve the performance of spin qubits for high-temperature operation, where the spin coherence is relatively short. The mechanism of this ultrafast $f_{\text{Rabi}}$ still needs further research, but this experimental phenomenon can at least prove that our system has a strong SOI.

We also extract the coherence time of this spin qubit in Section S6 of the supporting information, where the coherence time $T_2^* = 21.5$ ns. In Section S7 of the supporting information, we studied the influence of $\epsilon$ and $V_M$ on $T_2^*$. $T_2^*$ is generally on the order of 20 ns and has no obvious dependence on $\epsilon$ and $V_M$. This is probably because the large charge noise at this operation point plays the most important role in $T_2^*$. In our future research, we can perform a Hahn echo pulse sequence to extend the measured coherence time.[29] This can help us further study the influence of electric field parameters on coherence time. Combining the study of $f_{\text{Rabi}}$ and $T_2^*$, we can adjust the quality factor of the qubit by changing $\epsilon$ and $V_M$. In Section S8 of the supporting information, we show the dependence of quality factors on $\epsilon$ and $V_M$. It may help find the best operation point for the future study of spin qubits.

In conclusion, we fabricated a DQD based on a Ge HW. We define and manipulate a single-hole spin qubit with the mechanism of strong SOI in this material and measure the leakage current to determine whether PSB is lifted to obtain the state of the spin qubit. On the basis of manipulating and measuring the spin state, we focus on how $f_{\text{Rabi}}$ is affected by the electric field, including the detuning energy ($\epsilon$) and middle gate voltage ($V_{\text{M}}$). We have established a hole EDSR model and introduce the contribution of the excited state to $f_{\text{Rabi}}$. The fitting results show that $\epsilon$ and $V_{\text{M}}$ have obvious modulation effects on the SOI so as to affect $f_{\text{Rabi}}$. Additionally, $\epsilon$ can also modify the orbital states and thus affects $f_{\text{Rabi}}$. Moreover, we achieve an ultrafast $f_{\text{Rabi}} >$ 1.2 GHz, which is the highest value among semiconductor spin qubits. The above measurement results of $f_{\text{Rabi}}$ show the strong SOI in the system. Our work paves the way for research on scalable spin qubits.

**Supporting information**

The supporting information is available free of charge via the internet at http://pubs.acs.org.

Calculation process of the lever arm of plunger gates; simulation results of EDSR; theory of ultrafast EDSR of holes; specific fitting results of $f_{\text{Rabi}}$ as a function of electric field parameters; fitting results of $f_{\text{Rabi}}$ considering the possible relationship between tunnel coupling and gate voltage; decoherence time of the hole spin qubit; decoherence time as a function of electric field parameters; and quality factor of the spin qubit as a function of electric field parameters.

**Acknowledgments**

This work was supported by the National Natural Science Foundation of China (Grants No. 12074368, 92165207, 12034018 and 92265113), the Innovation Program for Quantum Science and Technology (Grant No. 2021ZD0302300), the Anhui Province Natural Science Foundation (Grants No. 2108085J03), and the USTC Tang Scholarship. P. H. acknowledges support from the National Natural Science Foundation of China (11904157), Shenzhen Science and Technology Program (No. KQTD20200820113010023), and Guangdong Provincial Key Laboratory (No.